\documentclass[%
 aip,
 jmp,%
 amsmath,amssymb,
preprint,
]{revtex4-1}

\usepackage{graphicx}
\usepackage{dcolumn}
\usepackage{bm}

\begin{document}


\title[]{Rotating Leaks in the Stadium Billiard}

\author{B. Appelbe}
\email{bappelbe@ic.ac.uk}
\affiliation{%
The Blackett Laboratory, Imperial College, London, SW7 2AZ, United Kingdom
}%

\date{\today}

\begin{abstract}
The open stadium billiard has a survival probability, $P(t)$, that depends on the rate of escape of particles through the leak. It is known that the decay of $P(t)$ is exponential early in time while for long times the decay follows a power law. In this work we investigate an open stadium billiard in which the leak is free to rotate around the boundary of the stadium at a constant velocity, $\omega$. It is found that $P(t)$ is very sensitive to $\omega$. For certain $\omega$ values $P(t)$ is purely exponential while for other values the power law behaviour at long times persists. We identify three ranges of $\omega$ values corresponding to three different responses of $P(t)$. It is shown that these variations in $P(t)$ are due to the interaction of the moving leak with Marginally Unstable Periodic Orbits (MUPOs).
\end{abstract}

\maketitle


\section{Introduction}\label{sec:1}
Chaotic billiards is a theory of dynamical systems that is popular due to its relatively simple formulation, rich dynamics and its application in areas such as mechanical processes, molecular dynamics and optics.\cite{Chernov_2006} Within this theory, the topic of open chaotic billiards, in which particles can escape from the billiard through a leak, has received much attention in recent years due to the presence of transient chaos. Transient chaos is chaotic motion with a finite lifetime that cannot be studied using asymptotic behaviour alone.\cite{Tel_2011, Tel_Chaos2015} Open billiards also have applications in a wide variety of physical phenomena including quantum chaos, astronomy and hydrodynamics.\cite{Altmann_RevModPhys2013}

The stadium billiard, first studied by Bunimovich, consists of two parallel straight edges with two semi-circular curves at each end.\cite{Bunimovich_1974,Bunimovich_1979} The closed stadium billiard exhibits strongly chaotic behaviour due to the defocusing mechanism that can occur when particles hit the curved boundary. Early studies of the open stadium billiard, in which a small leak was placed on the boundary of an otherwise closed stadium billiard (see fig. \ref{f:1}), revealed interesting results for the fraction of initial trajectories remaining within the billiard as a function of time.\cite{Alt_PRE1995, DUMONT_CPL1992} This fraction is referred to as the survival probability, $P(t)$. It was shown that $P(t)$ decreases exponentially at early times ($P(t)\sim\exp\left(-\kappa t\right)$). However, at longer times $P(t)$ asymptotically approaches a power-law ($P(t)\sim t^{-z}$). This power-law response is referred to as the formation of an algebraic tail of the survival probability.

It is known that the algebraic tail is due to the presence of ``sticky regions'' in the phase space of the billiard.\cite{Armstead_PhysD2004} An invariant set in phase space is deemed sticky if orbits approaching the invariant set remain close to the invariant set for a long time.\cite{Bunimovich_Chaos2012, Dettmann_arXiv2016} In the stadium billiard, the ``sticky'' invariant set consists of orbits that collide with the straight edge orthogonally and so bounce back and forth between the straight edges. Orbits colliding with the straight edge at an angle close to orthogonal are referred to as Marginally Unstable Periodic Orbits (MUPOs).\cite{Dettmann_PhysD2009} These MUPOs move slowly through the phase space and so can take much longer than a chaotic orbit to reach some regions of phase space, resulting in the algebraic tail of $P(t)$.

In this work we investigate the behaviour of $P(t)$ when the leak rotates at constant velocity around the boundary of the stadium billiard. Our intuition tells us that a rotating leak should eliminate the algebraic tail since the leak will move to regions of phase space in which MUPOs exist rather than MUPOs having to reach the location of a leak fixed in position. However, our results show that the picture is more complicated than this. There are indeed values of rotation velocity, $\omega$, for which the decay of $P(t)$ is purely exponential but we also find values for which the algebraic tail persists. Finally, we find values of $\omega$ for which an algebraic tail of $P(t)$ begins to form at intermediate times but exponential decay resumes at later times.

There has been a wide variety of studies of open billiards,\cite{Altmann_RevModPhys2013} including the effects of noise perturbations on trajectories,\cite{Altmann_Chaos2012} multiple leaks\cite{Bunimovich_PRL2005,Dettmann_PRE2011}, vibrating boundaries\cite{Leonel_PLA2012} and leak size.\cite{Bunimovich_EPL2007} However, it appears that there is only one previous work on a billiard with a moving leak.\cite{Hansen_arXiv2016} In this work by Hansen et al., each particle was simulated separately and the leak location changed if the particle escaped or after $5$ collisions with the boundary. Thus, the leak dynamics were dependent on the particle dynamics and were different for each particle trajectory. In our work, we study a leak that is completely independent of the particle dynamics.

The contents of this paper are as follows. In section \ref{sec:2} we introduce the open stadium billiard and describe the set-up of our numerical model. In section \ref{sec:3} we described results for a stationary leak. This provides a useful basis for interpreting the results from a rotating leak. Section \ref{sec:4} describes the results for the rotating leaks, divided into subsections for low, medium and high rotation velocities. Section \ref{sec:5} discusses the effects of varying the size of the rotating leak and, finally, section \ref{sec:5} summarizes our results and contains some concluding remarks.

\section{The Bunimovich stadium and simulation set-up}\label{sec:2}
\begin{figure}
\begin{center}%
\includegraphics*[width=0.9\columnwidth]{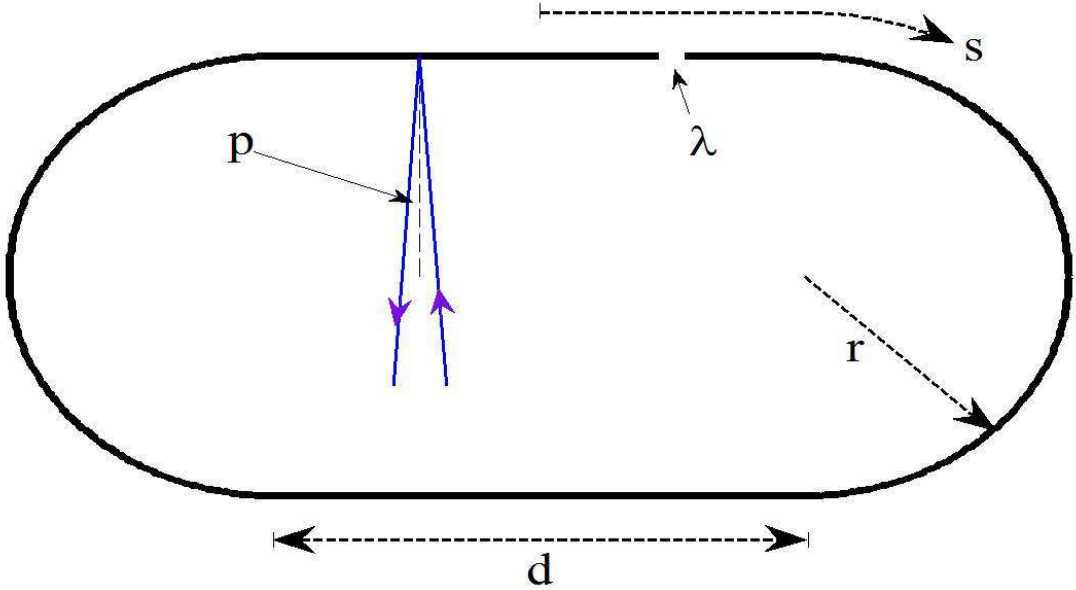}%
\vspace{-1em}
\end{center}
\caption[]{The Bunimovich stadium with a leak of width $\lambda$. When a particle trajectory collides with a straight edge with a $p$ value close to $0$ we say that the particle is in a Marginally Unstable Periodic Orbit (MUPO). Such particles will progress slowly along the straight edges.} \label{f:1}
\end{figure}

A sketch of the Bunimovich stadium is shown in fig. \ref{f:1}. The stadium consists of two straight, parallel edges of length $d$, joined to semi-circles of radius $r$ at each end. Particles move in straight lines in the stadium, have a constant velocity and are non-interacting (they cannot collide with each other). When a particle reaches the stadium boundary it undergoes a specular reflection, meaning that the incoming and outgoing trajectories make the same angle with the normal to the boundary. The open stadium billiard contains a leak, a portion of the boundary through which particles escape rather than being reflected.

Birkhoff coordinates $\textbf{x}=\left(s,p\right)$, where $s\,\epsilon\,\left[0,1\right]$ and $p\,\epsilon\,\left[-1,1\right]$, are used to define particle collisions. Here, $s$ is particle location on boundary measured clockwise from the midpoint of the top edge of the stadium and $p = \sin\theta$, where $\theta$ is the angle between the particle trajectory after collision and the inward pointing normal to the boundary. The value $\theta = \pi/2\,\left(-\pi/2\right)$ corresponds to a particle moving along the boundary in the direction of increasing (decreasing) $s$.

We define a leak in the stadium billiard as the range of $s$ values $\left[s_L(t)-\lambda/2,s_L(t)+\lambda/2\right]$ where $s_L(t)$ represents the location of the centre of the leak at time $t$ and $\lambda$ is the width of the leak (constant in time). If the $s$ coordinate of a particle during any collision lies within the leak then that particle does not undergo a reflection but instead escapes from the system and its trajectory is no longer tracked. We use $\omega$ to denote the rotation velocity of the leak in units of cycles per unit time (i.e. $\omega = 1$ results in one complete rotation of the leak around the stadium boundary in one unit of time).

For our calculations we use $d = 2$, $r = 1$, $\lambda = 0.01$ unless otherwise stated. For simulations with rotating leaks the leak is initialised at $s_L(0) = 0$ and moves in the $+s$ direction. In each simulation $10^{10}$ particles are initialised, uniformly distributed in $\left(s,p\right)$, with a particle velocity $v=1$. Each particle is created at a randomly chosen initial time between $t=0$ and $t=4$. This random initial time is used to avoid correlations in the collision times of nearby orbits. Finally, we note that several tests were carried out in which particles were initialised uniformly in the 3-dimensional billiard-flow phase space. These produced similar results to the uniform initialization in $\left(s,p\right)$ suggesting that the results we report are insensitive to the initial conditions used.

\section{Stationary Leaks}\label{sec:3}
\begin{figure}
\begin{center}%
\includegraphics*[width=0.99\columnwidth]{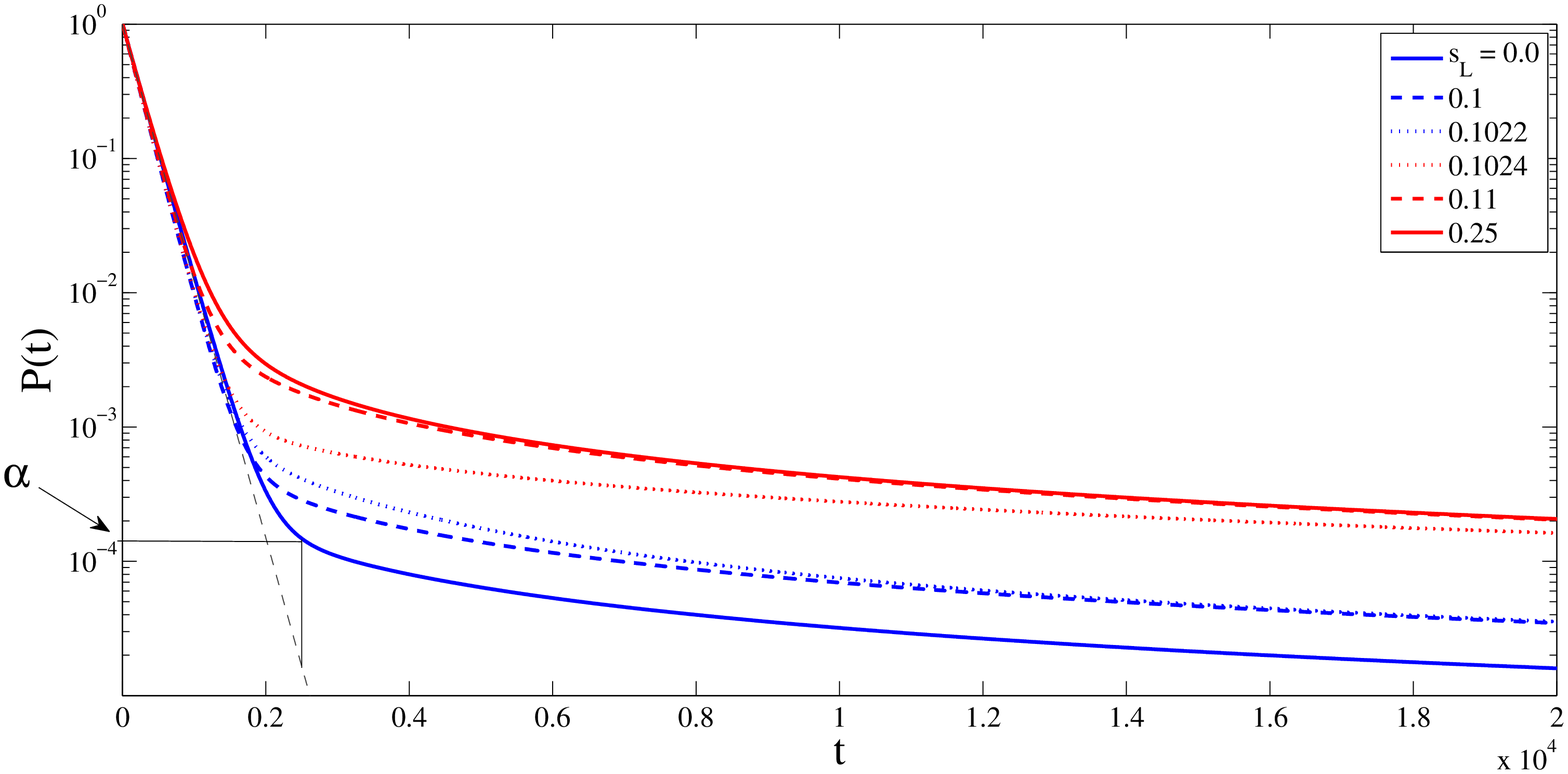}%
\vspace{-1em}
\end{center}
\caption[]{Survival probability $P(t)$ as a function of time, $t$, for a stationary leak located at $s_L$. The value of $\alpha$ for $s_L = 0.0$ is indicated.} \label{f:2}
\end{figure}

We begin our presentation of results with the case of a leak that is fixed in position ($\omega = 0$). In fig. \ref{f:2} we plot the survival probability for a number of fixed leak locations on both the straight and circular portions of the boundary. Due to the symmetry of the stadium we need only simulate the range of locations $s_L\,\epsilon\,\left[0,0.25\right]$.

The transition from exponential to algebraic decay is clear in all of the results in fig. \ref{f:2}. For early times ($t<\, \sim500$) the survival probability is well fitted by $\exp\left(-\kappa t\right)$. Measured values of the exponential decay constant $\kappa$ are given in table \ref{t:1}. It is also clear from fig. \ref{f:2} that the fraction of particles in the algebraic tail increases with increasing $s_L$. To quantify this we introduce the parameter $\alpha$. We define $\alpha$ to be the value of the survival probability when the survival probability exceeds the extrapolated purely exponential decay by an order of magnitude. Measured values of $\alpha$ are given in table \ref{t:1}.

\begin{table}\centering
\caption[My table caption]{Exponential decay coefficients, $\kappa$, and algebraic tail weightings, $\alpha$, for stationary leaks at locations $s_L$.}\label{t:1}
\begin{footnotesize}
\begin{tabular}{l l l l l }
\hline
\hline
$s_L$ & & $\kappa$ & & $\alpha$ \\
\hline
& & & & \\
$0$  & & $4.402\times 10^{-3}$  & & $1.4\times 10^{-4}$ \\
$0.1$  & & $4.748\times 10^{-3}$  & & $3.6\times 10^{-4}$  \\
$0.1022$  & & $4.752\times 10^{-3}$  & & $5.7\times 10^{-4}$  \\
$0.1024$  & & $4.750\times 10^{-3}$  & & $9.5\times 10^{-4}$ \\
$0.11$  & & $4.683\times 10^{-3}$  & & $2.9\times 10^{-3}$ \\
$0.25$  & & $4.382\times 10^{-3}$  & & $3.4\times 10^{-3}$ \\
& & & & \\
\hline
\hline
\end{tabular}
\end{footnotesize}
\end{table}

\begin{figure}
\begin{center}%
\includegraphics*[width=0.99\columnwidth]{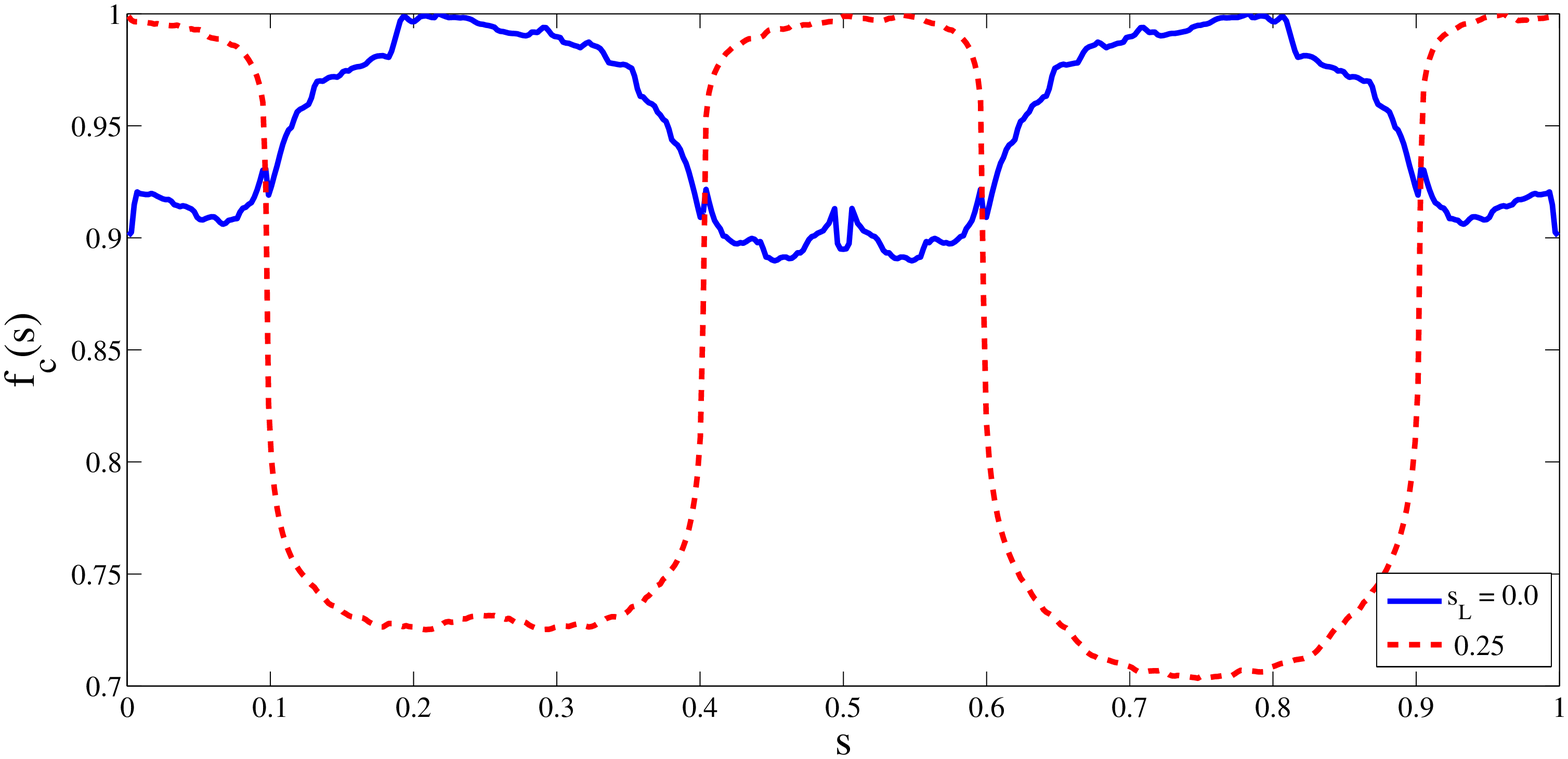}%
\hfil%
\includegraphics*[width=0.99\columnwidth]{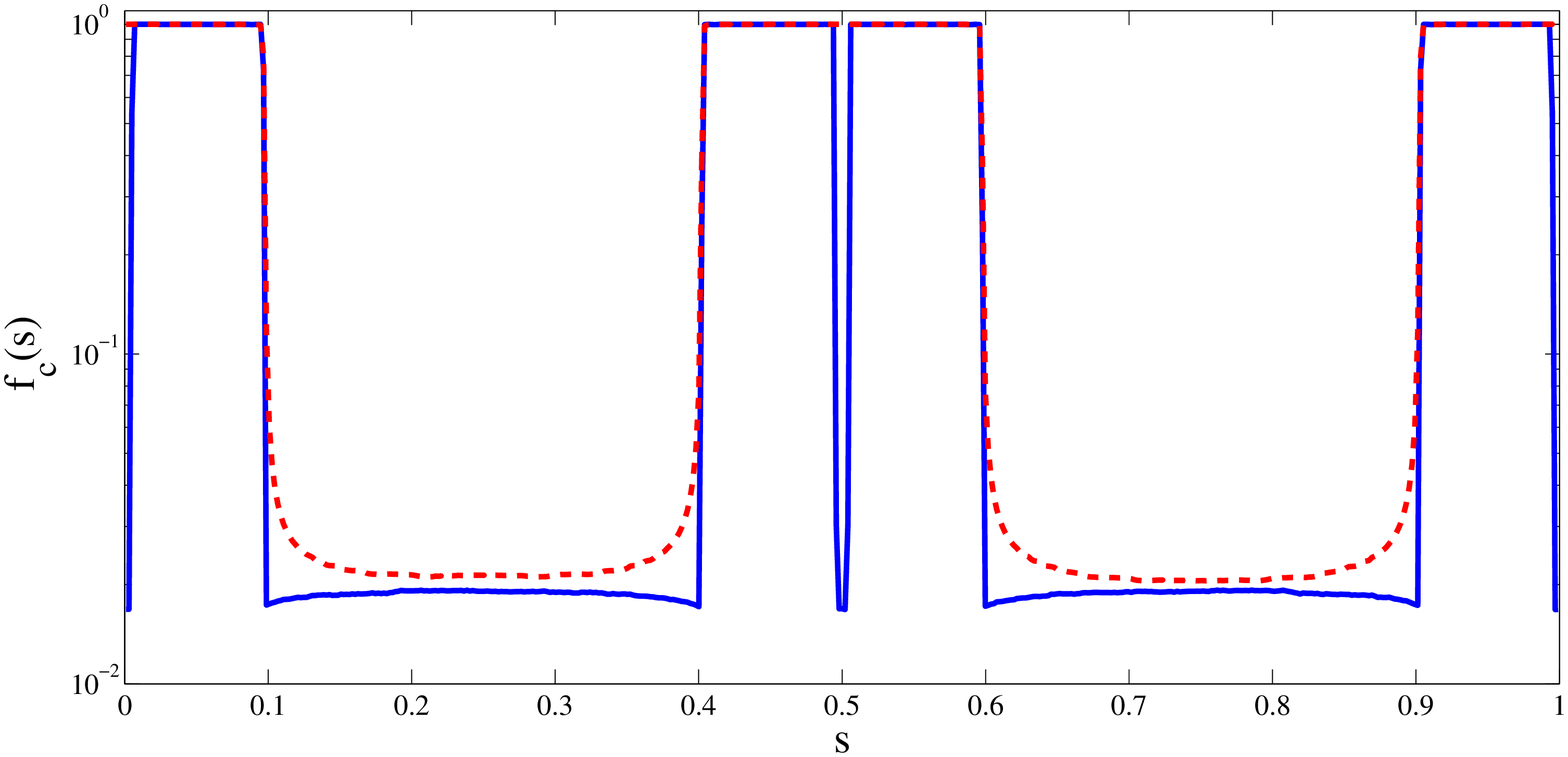}%
\vspace{-1em}
\end{center}
\caption[]{Top: The normalized distribution of collisions, $f_c\left(s\right)$, as a function of $s$ for simulations with a stationary leak at $s_L = 0.0$ and $s_L = 0.25$ for times $0\leq t\leq 2\times10^{4}$. Bottom: The same data for times $2\times10^{3} \leq t\leq 2\times10^{4}$. Note that the top diagram has a linear vertical scale while the bottom diagram has a logarithmic vertical scale.} \label{f:3}
\end{figure}

The major contribution made by MUPOs to the formation of the algebraic tail can be seen by comparing the two diagrams in fig. \ref{f:3}. In these diagrams, $f_c\left(s\right)$ is plotted as a function of the $s$ coordinate. The quantity $f_c\left(s\right)$ represents the number of collisions of particles at a particular location on the boundary. The top diagram in fig. \ref{f:3} shows the results for $s_L=0.0$ and $0.25$ for the full simulation time. It is clear that $f_c\left(s\right)$ is slightly greater in regions of $s$ corresponding to the straight boundary of the stadium for $s_L=0.25$. The opposite is true for $s_L=0.0$. However, in both cases the differences between minima and maxima of $f_c\left(s\right)$ is relatively small.

The bottom diagram of fig. \ref{f:3} shows $f_c\left(s\right)$ when only collisions occurring for $t\geq 2\times10^3$ are considered. This value of $t$ is chosen such that we are calculating $f_c\left(s\right)$ for the algebraic tail only. Here, $f_c\left(s\right)$ is more than an order of magnitude larger for regions corresponding to the straight boundary than the curved boundary for both $s_L=0.0$ and $0.25$. This is because the majority of orbits in the algebraic tail are MUPOs, with particles bouncing back and forth between the straight edge boundaries. The value of $\alpha$ is larger for $s_L = 0.25$ than for $s_L = 0.0$. This is because, in order to reach the leak at $s_L=0.25$, particles in MUPOs must first reach the curved region of the boundary and then decay from the MUPO to a chaotic orbit. This switching of trajectories from regular to chaotic motion (and vice versa) is referred to as intermittency and is a well-known property of stadium billiards.\cite{Dettmann_Chaos2012} In contrast for a leak on the straight portion of the boundary (such as $s_L = 0.0$) particles can reach the leak whilst in a MUPO.

Finally, fig. \ref{f:2} shows an abrupt transition in the algebraic tail between leak locations $s_L = 0.1022$ and $0.1024$. The point at which the straight boundary joins the curved boundary is $s=0.09725$ and, since we are using $\lambda = 0.01$, a value of $s_L = 0.1022$ means that a small fraction of the leak still lies on the straight boundary edge but for a value of $s_L = 0.1024$ the leak lies entirely on the curved boundary. Therefore, for $s_L=0.1022$ particles can reach the leak whilst in a MUPO but for $s_L=0.1024$ particles need to first switch from a MUPO to a more chaotic orbit.

\section{Rotating Leaks}\label{sec:4}
The variations in the survival probability for stationary leaks at different locations, as described in section \ref{sec:3}, suggests that a leak which is moving along the boundary as particle trajectories evolve could result in new behaviour of the survival probability. In particular, our intuition suggests that a moving leak may cause the survival probability to decay faster since the leak will move to regions of phase space in which MUPOs are located. However, as we illustrate in the next three subsections, the actual behaviour is more complicated than this. Each subsection corresponds to a different range of rotation velocities which each display different characteristics in the dependence of survival probability on rotation velocity.

\subsection{Rotating Leaks I: Low-$\omega$ Behaviour}\label{sec:4.1}
The first regime that we study is for small values of $\omega$ such that the timescale required for the leak to complete one full cycle around the stadium boundary is greater than the timescale for the formation of the algebraic tail of the survival probability.

\begin{figure}
\begin{center}%
\includegraphics*[width=0.99\columnwidth]{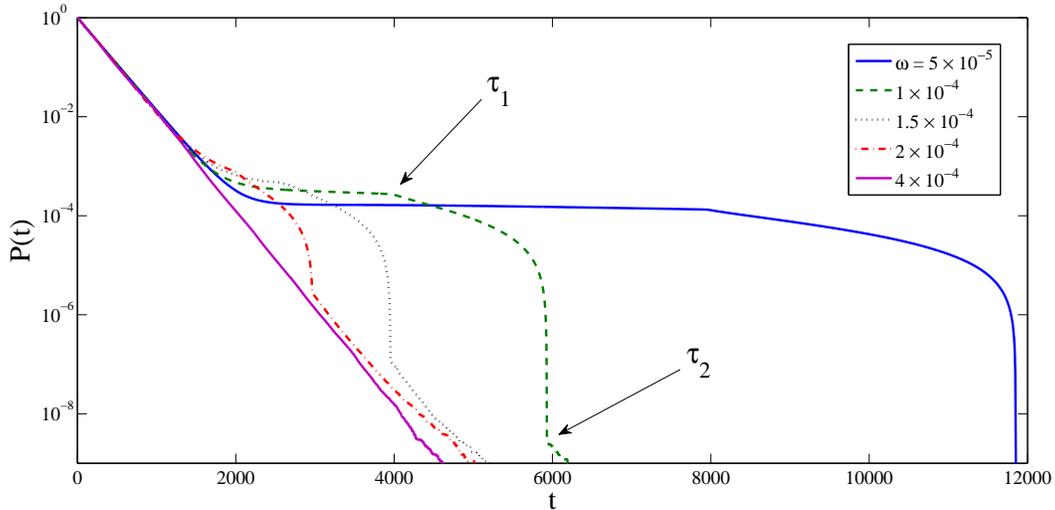}%
\vspace{-1em}
\end{center}
\caption[]{Survival probabilities, $P(t)$, for a number of rotation velocities in the low-$\omega$ regime.} \label{f:4}
\end{figure}

Results for $P(t)$ for a number of simulations in the low-$\omega$ regime are shown in in fig. \ref{f:4}. This figure clearly illustrates the characteristic behaviour of $P(t)$ in this regime: an algebraic tail is formed at intermediate times but this is later eliminated with exponential decay of $P(t)$ resuming. The range of $\omega$ which results in low-$\omega$ behaviour of $P(t)$ is $0< \omega< \sim4\times 10^{-4}$.

To understand this behaviour we take the example of $\omega = 1\times 10^{-4}$. The leak starts at $s_L = 0$ and reaches the transition from the straight to curved boundary at $t = 972.5$. This time is before the algebraic tail has become apparent in $P(t)$ and the exponential decay coefficient is approximately $\kappa = 4.393\times10^{-3}$. As the leak moves around the curved boundary the algebraic tail develops with $\alpha = 3.7\times 10^{-4}$. As illustrated in fig. \ref{f:4}, at a time marked $\tau_1 \approx 3980$ there is a sharp change in $P(t)$ with a large increase in the rate of decay. The time $\tau_1$ corresponds to when the leading edge of the leak is moving from the curved boundary to the lower straight boundary. Therefore, the increased rate of decay of $P(t)$ at times greater than $\tau_1$ is due to the loss of MUPOs as the leak moves along the lower straight boundary. As the leak moves further along the curved boundary the rate of decay increases even more. This is because a larger number of MUPOs exist in the left half of the straight boundaries than the right since some of the MUPOs on the right will have leaked early in time when the leak was moving along the upper straight boundary. This increased decay continues until $\tau_2 \approx 5930$, which corresponds to when the leading edge of the leak reaches the left curved boundary. After this time, the leak moves along the curved boundary and exponential decay resumes with a decay coefficient similar to that of the early time.

The behaviour of $P(t)$ is similar for other values of $\omega$ in the low-$\omega$ regime with the corresponding values of $\tau_1$ and $\tau_2$ being larger (smaller) for slower (faster) moving leaks. As $\omega$ increases $\tau_1$ and $\tau_2$ move closer to the region of early time exponential decay. Eventually, $\omega$ is sufficiently large such that the algebraic tail does not have time to form before the leak starts moving along the lower boundary and sweeping up the MUPOs. The decay of $P(t)$ becomes purely exponential. We find that this occurs at $\omega \approx 4\times 10^{-4}$. Therefore, we take this value to be the upper limit of the low-$\omega$ regime.

\subsection{Rotating Leaks II: Mid-$\omega$ Behaviour}\label{sec:4.2}
The mid-$\omega$ regime is characterised by survival probability having purely exponential decay for all values of $\omega$. We find no evidence of an algebraic tail of $P(t)$ in this regime. Our simulations suggest that the range of the mid-$\omega$ regime is $\sim 4\times 10^{-4}\leq\omega < \sim0.01$.

Physically, the purely exponential decay in the mid-$\omega$ regime can be understood in terms of the interaction between the moving leak and MUPOs. For $\omega$ values in the mid-$\omega$ range, the leak is moving fast enough to sweep up the MUPOs before an algebraic tail can form (this is in contrast to the low-$\omega$ regime). However, the leak is also moving sufficiently slowly such that it does not jump beyond MUPOs as it moves around the boundary. For example, orbits in the ``sticky'' invariant set are bouncing back and forth between the same two points on the straight boundaries with a period of $4$. Therefore, for $\lambda = 0.01$, if we have $\omega\leq 2.5\times10^{-3}$ these orbits are certain to intersect the leak as it moves along the straight boundary. For larger values of $\omega$ the leak can move beyond the reflection point of the orbit on the boundary while the particle is in transit in the stadium. MUPOs with a value of coordinate $p$ close to $0$ will similarly be unable to avoid a slow moving leak. Some MUPOs with a sufficiently large $p$ can avoid the leak because of the displacement along the boundary with each reflection. However, it appears that there is not a sufficiently large population of these orbits to alter the exponential decay behaviour of $P(t)$ for $\omega < \sim 0.01$.

For all values of $\omega$ studied in the mid-$\omega$ regime it is found that the exponential decay coefficient lies in the range $\kappa\,\epsilon\,\left[4.5\times10^{-3},4.7\times10^{-3}\right]$. This is similar to the values of $\kappa$ that were obtained for stationary leaks and leaks in the low-$\omega$ and high-$\omega$ regimes. Finally, we do not rule out the formation of an algebraic tail with a very low value of $\alpha$ due to the presence of long-lived orbits in the stadium. However, our studies suggest that this tail will have $\alpha\leq 10^{-9}$.

\subsection{Rotating Leaks III: High-$\omega$ Behaviour}\label{sec:4.3}
For $\omega\geq \sim0.01$ new behaviour of $P(t)$ with changing $\omega$ begins to emerge. We refer to this as the regime of high-$\omega$ behaviour. The relationship between $\omega$ and $P(t)$ appears to be significantly more complex for the high-$\omega$ regime than for both the low-$\omega$ and mid-$\omega$ regimes.

\begin{figure}
\begin{center}%
\includegraphics*[width=0.99\columnwidth]{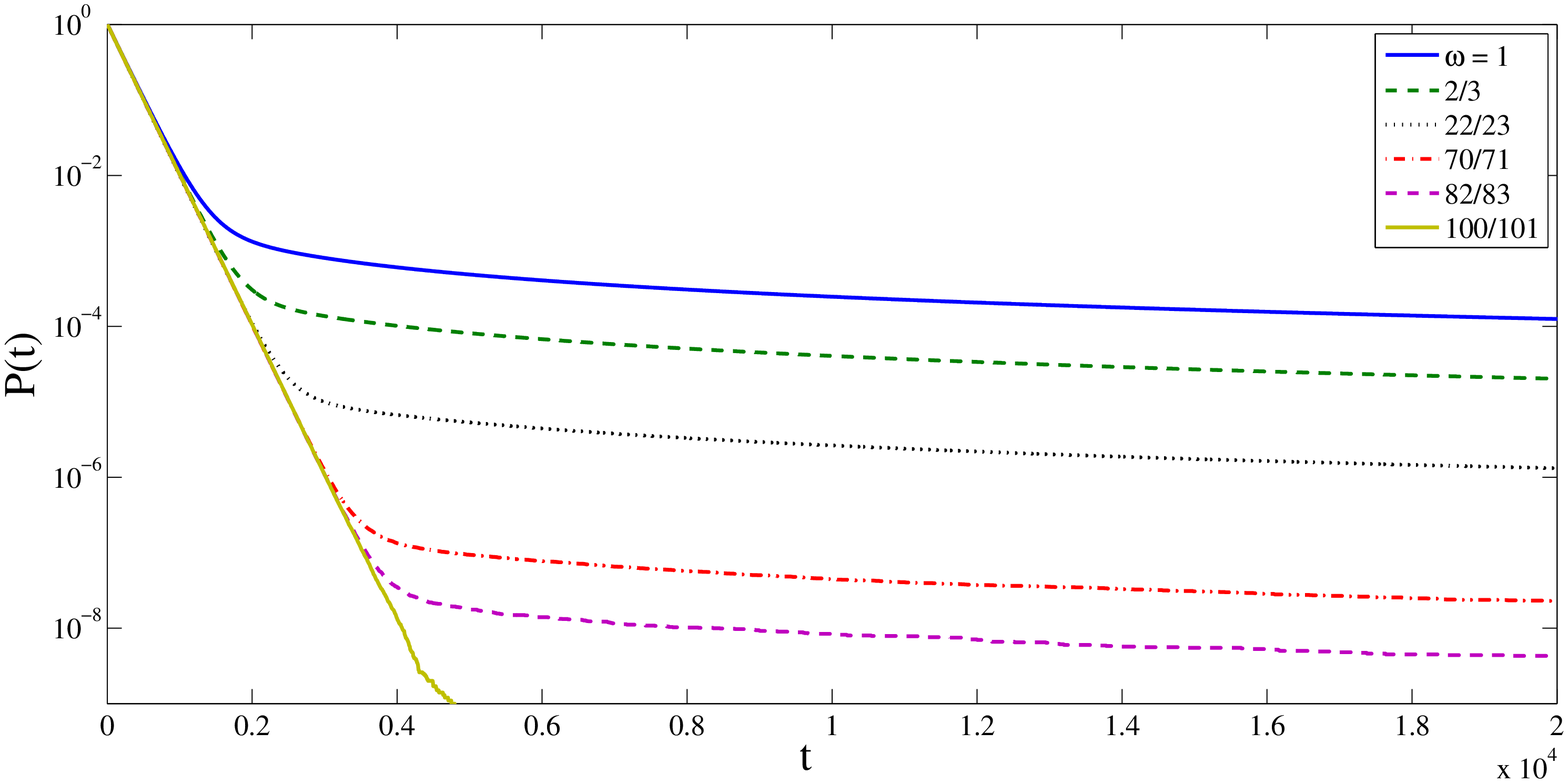}%
\hfil%
\includegraphics*[width=0.99\columnwidth]{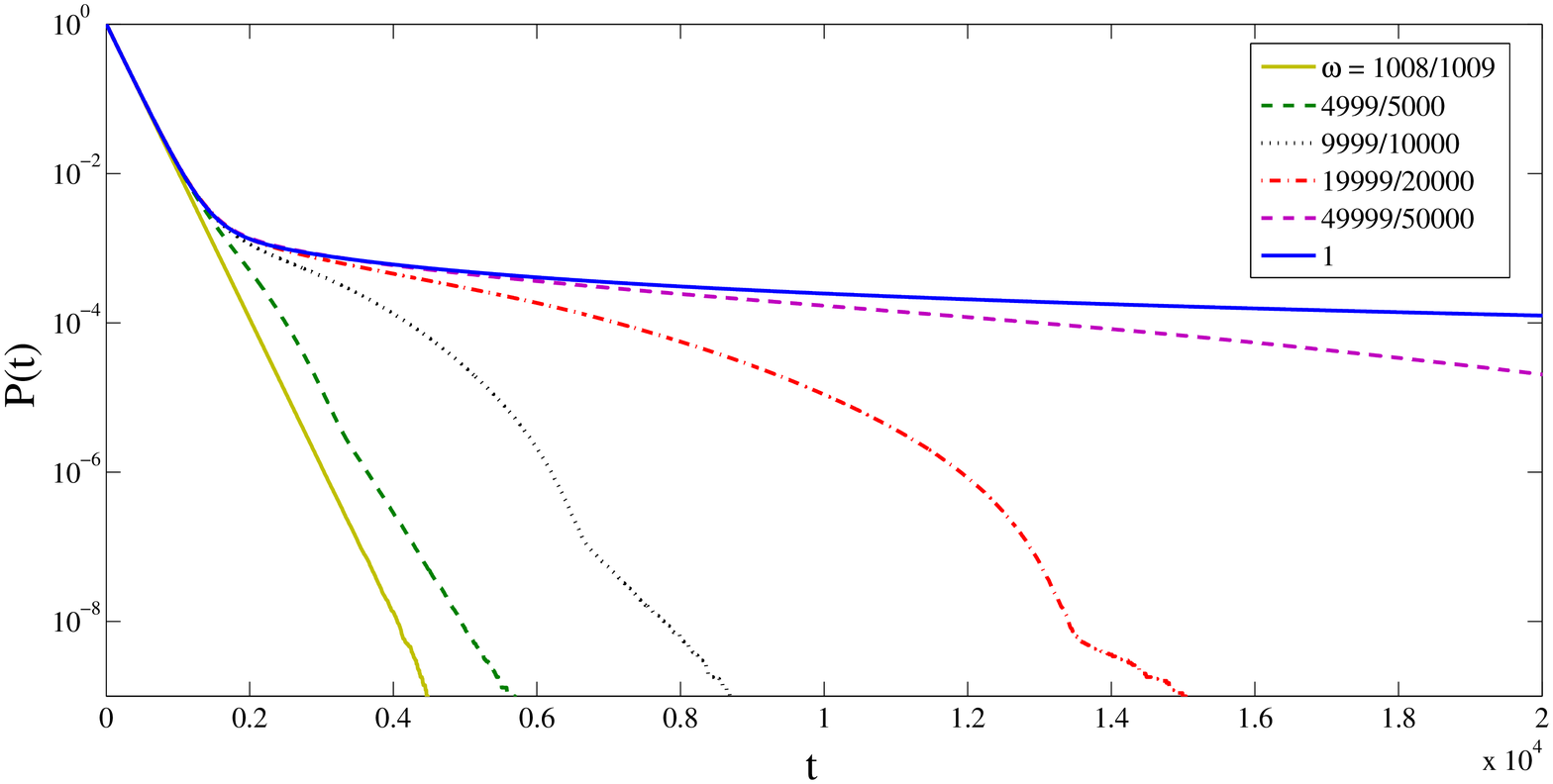}%
\vspace{-1em}
\end{center}
\caption[]{Top: Survival probabilities for $\omega=1$ and a number of successively closer approximations to $1$ in which an algebraic tail is evident with decreasing $\alpha$. Bottom: Survival probabilities for successively closer approximations to $\omega = 1$ in which $P(t)$ asymptotically approaches the $\omega = 1$ curve.} \label{f:5}
\end{figure}

We first present results for rational values of rotation velocity, $\omega = \frac{p}{q}$ where $p,q$ are coprime integers. The case of irrational $\omega$ is discussed at the end of the section. Figure \ref{f:5} illustrates the dependence of $P(t)$ on $\omega$ for a number of rational values of $\omega\,\left(=\frac{q-1}{q}\right)$ that are successively closer approximations $\omega = 1$. There are two clear trends in $P(t)$. First, for ``smaller'' values of denominator $q$, $P(t)$ contains an algebraic tail with a value of $\alpha$ that decreases rapidly with increasing $q$. This trend is illustrated in the top diagram of fig. \ref{f:5} with the corresponding values of $\kappa$ and $\alpha$ given in table \ref{t:2}. The number of particles that we have simulated means that we are unable to tell if the algebraic tail persists for values of $q>\sim 100$. However, as we continue to increase $q$ a second trend becomes apparent for $q>\sim 1000$. This is illustrated in the bottom diagram of fig. \ref{f:5}. Here we see that $P(t)$ has a purely exponential decay for $\omega = \frac{1008}{1009}$ but as the rational approximations get closer and closer to $\omega = 1$ the curve $P(t)$ asymptotically approaches the curve corresponding to $\omega = 1$.

\begin{table}\centering
\caption[My table caption]{Exponential decay coefficients, $\kappa$, and algebraic tail weightings, $\alpha$, for leaks rotating with rotation velocity $\omega$.}\label{t:2}
\begin{footnotesize}
\begin{tabular}{l l l l l }
\hline
\hline
$\omega$ & & $\kappa$ & & $\alpha$ \\
\hline
& & & & \\
$1$  & & $4.533\times 10^{-3}$  & & $1.4\times 10^{-3}$ \\
& & & & \\
$\frac{2}{3}$  & & $4.609\times 10^{-3}$  & & $1.9\times 10^{-4}$  \\
& & & & \\
$\frac{22}{23}$  & & $4.633\times 10^{-3}$  & & $1.0\times 10^{-5}$  \\
& & & & \\
$\frac{70}{71}$  & & $4.606\times 10^{-3}$  & & $1.4\times 10^{-7}$ \\
 & & & & \\
$\frac{82}{83}$  & & $4.604\times 10^{-3}$  & & $2.5\times 10^{-8}$ \\
& & & & \\
$\frac{100}{101}$  & & $4.600\times 10^{-3}$  & & $<1.0 \times 10^{-9}$ \\
& & & & \\
\hline
\hline
\end{tabular}
\end{footnotesize}
\end{table}

Our numerical simulations indicate that similar trends in $P(t)$ are present for other rational values of $\omega$. For small values of the denominator $q$ an algebraic tail is present. As $q$ increases the $\alpha$ value of the algebraic tail decreases. This trend continues as $q$ increases but is interrupted if $\frac{p}{q}$ becomes sufficiently close to a rational number with denominator less than $q$ (in our example in fig. \ref{f:5} this occurs near $\omega = \frac{1008}{1009}$ which is sufficiently close to $\omega=1$).

\begin{figure}
\begin{center}%
\includegraphics*[width=0.99\columnwidth]{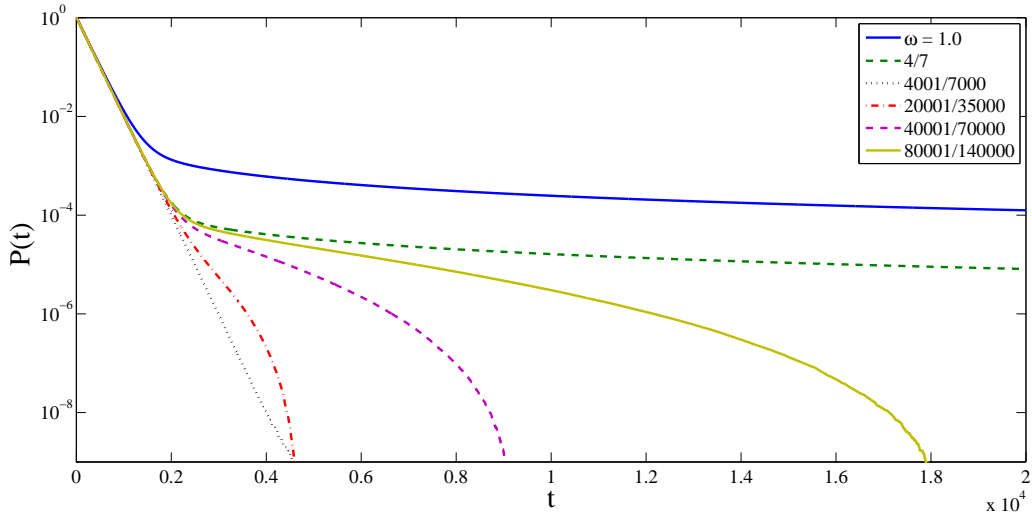}%
\vspace{-1em}
\end{center}
\caption[]{Survival probabilities for $\omega=1,\frac{4}{7}$ and a number of rational approximations to $\frac{4}{7}$. The $\alpha$ value for $\omega=\frac{4}{7}$ is $7.1\times 10^{-5}$.} \label{f:6}
\end{figure}

Another example illustrating these trends is shown in fig. \ref{f:6}. Here, the decay of $P(t)$ appears to be purely exponential for $\frac{p}{q} = \frac{4001}{7000}$ but closer approximations to $\frac{4}{7}$ cause $P(t)$ to asymptotically recover its algebraic tail behaviour. We also note that the survival probability $P(t)$ for $\omega = \frac{p}{7}$ where $p\,\epsilon\,\left[1,6\right]$ is almost identical to that shown for $\omega = \frac{4}{7}$. In the high-$\omega$ regime in general, we find that for values of $\omega$ in which $P(t)$ has an algebraic tail, changing the value of the numerator $p$ does not change the $\alpha$ value of the algebraic tail as long as $\frac{p}{q}$ does not too closely approximate a rational $\omega$ with lower denominator value $q$.

The complex behaviour of $P(t)$ described above can be understood by assuming that MUPOs are responsible for the algebraic tail. If the moving leak can intersect the MUPOs then an algebraic tail is unlikely to form. For a given MUPO with coordinate value $p$ the displacement of the orbit along the straight boundary, $\Delta x$, and time, $t_c$, between consecutive collisions is
\begin{eqnarray}\label{e:1}
 \Delta x &=& \frac{2rp}{\sqrt{1-p^2}},\\
t_c &=& \frac{2r}{\sqrt{1-p^2}}.
\end{eqnarray}
Here, we are assuming that particle velocity is $1$. The number of reflections, $N$, required for the particle to move the maximum distance possible along the straight boundary, $d$, is given by
\begin{equation}\label{e:2}
N=\frac{d}{\Delta x} = \frac{d\sqrt{1-p^2}}{2rp}.
\end{equation}
Now since $p$ is small we can let
\begin{equation}\label{e:3}
1+\delta = \frac{1}{\sqrt{1-p^2}},
\end{equation}
where $\delta$ is a small parameter. This results in
\begin{eqnarray}\label{e:4}
 t_c &=& 2r\left(1+\delta\right),\\
 N &\sim& \frac{d}{2r}\frac{1}{\sqrt{\delta}}.
\end{eqnarray}
We can now calculate the location of the leak at the time of each reflection of the MUPO (assuming $s_L = 0$ when the orbit begins) as
\begin{equation}\label{e:5}
s_L\left(nt_c\right) = nt_c\omega-\lfloor nt_c\omega\rfloor\quad\forall\quad n\leq N.
\end{equation}
We can use this to calculate how well the leak covers the boundary during the $N$ reflections of the MUPO. We use $f_L\left(s\right)$ to denote this covering. This quantity represents the number of times during the $N$ reflections that the leak (which covers the range $\left[s_L\left(nt_c\right)-\lambda/2,s_L\left(nt_c\right)+\lambda/2\right]$) covers a coordinate value $s$. It gives a distribution of the leak locations that occur when the MUPO strikes the boundary.

\begin{figure}
\begin{center}%
\includegraphics*[width=0.99\columnwidth]{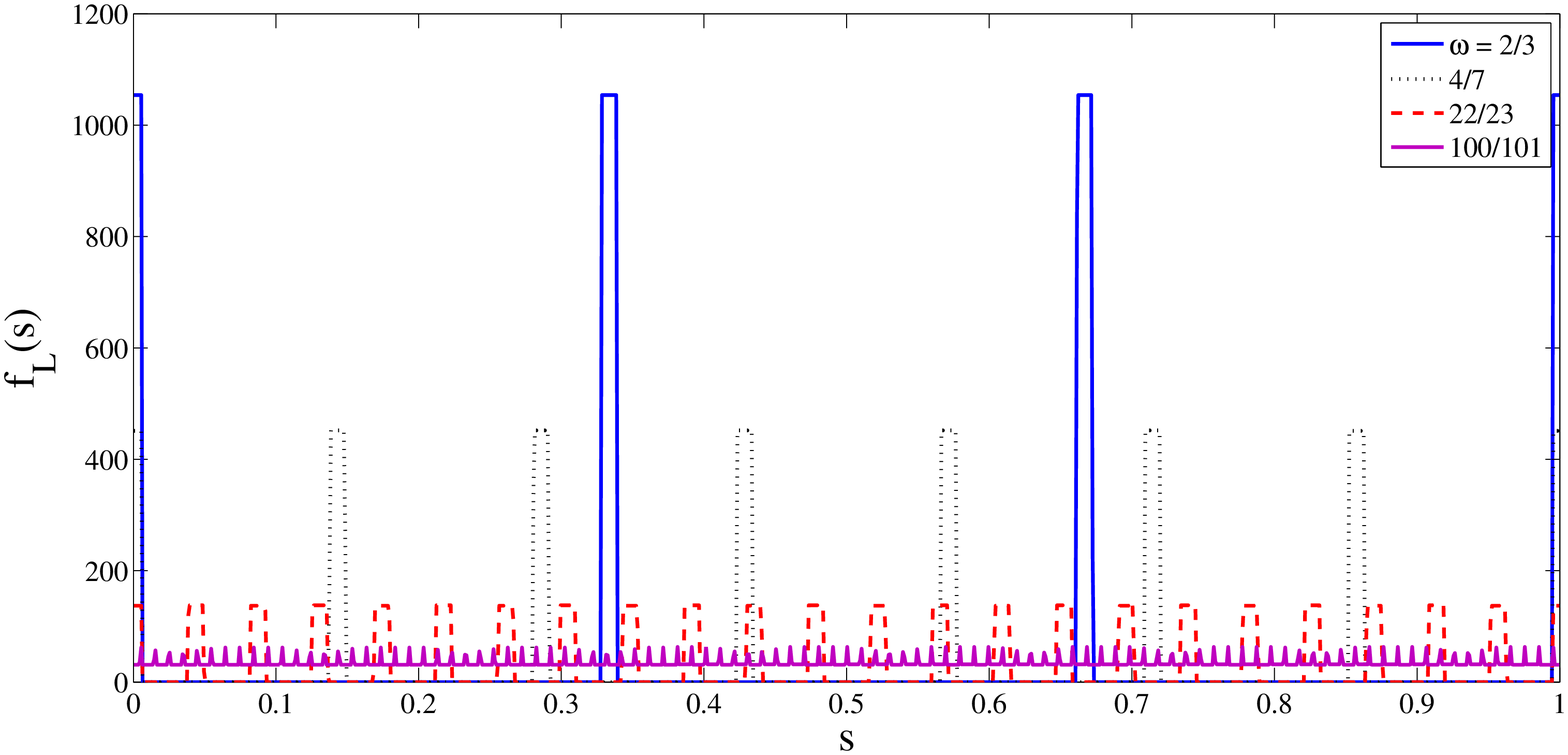}%
\hfil%
\includegraphics*[width=0.99\columnwidth]{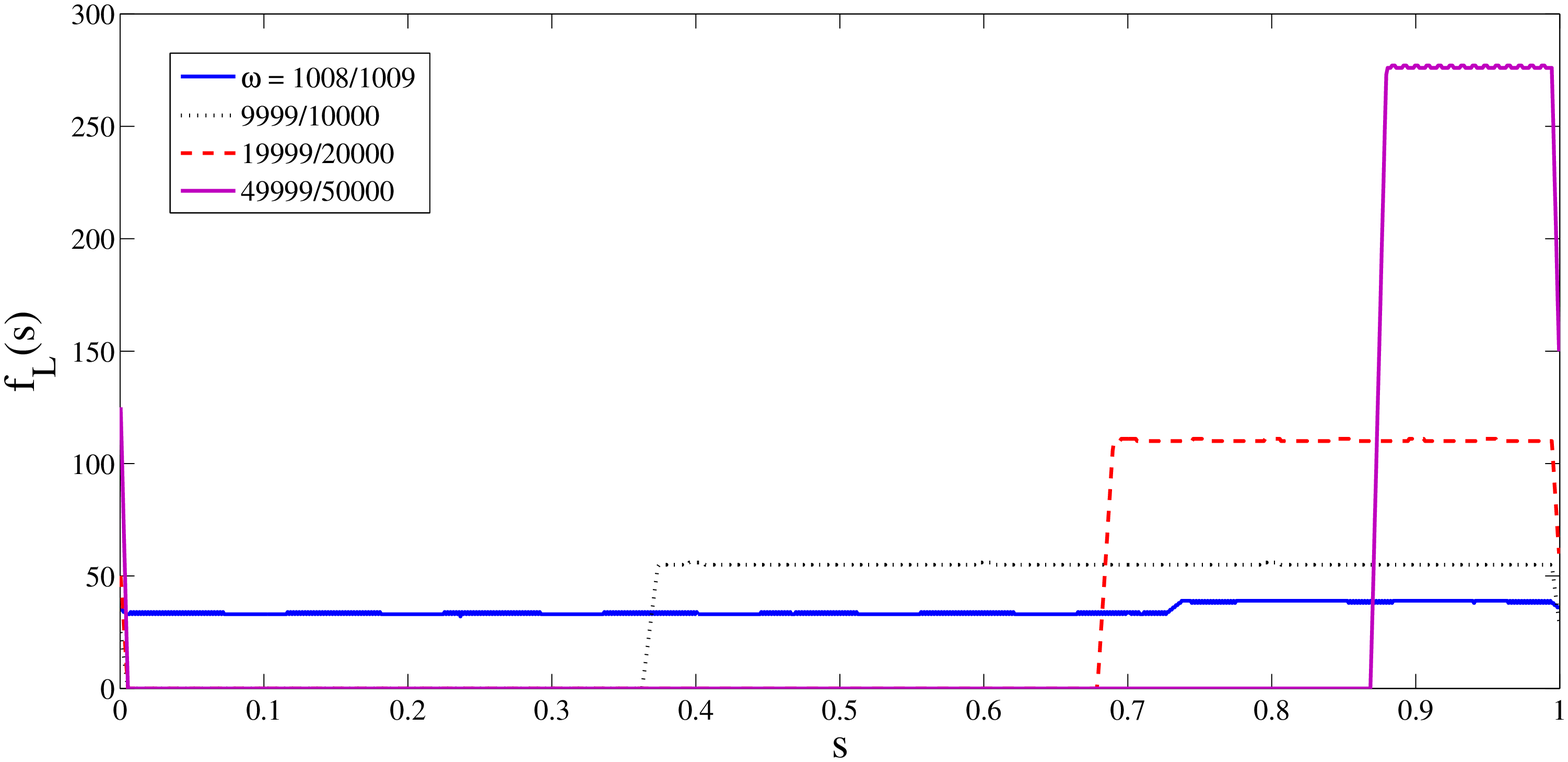}%
\vspace{-1em}
\end{center}
\caption[]{Top: The covering, $f_L(s)$, of the leak of the domain, $s$, for a number of $\omega$ values that result in an algebraic tail in $P(t)$ (cf. fig. \ref{f:5}). Bottom: The same quantity for $\omega$ values that asymptotically approach an algebraic tail of $P(t)$ as $\omega$ increases. We have used $\delta=10^{-7}$ for both diagrams. Similar results are obtained when the value of $\delta$ is varied.} \label{f:7}
\end{figure}

 If we calculate $f_L\left(s\right)$ for various values of $\omega$ then a clear trend emerges that matches the observed trends in $P(t)$ in the high-$\omega$ regime, as illustrated in fig. \ref{f:7}. The less uniform the covering of the domain, $s$, then the larger the $\alpha$ value of the algebraic tail. For example, for $\omega=2/3$, during the evolution of a MUPO the leak is near one of three locations ($s_L = 0,1/3,2/3$) each time the particle strikes the boundary. However, for $\omega = 100/101$ the covering $f_L\left(s\right)$ is much more uniformly distributed and this corresponds to $P(t)$ (shown in fig. \ref{f:5}) in which the algebraic tail has become undetectable.

Intuitively, we can understand this result as follows. A more uniform covering means that the leak is in a different location every time the MUPO strikes the boundary. Since the MUPOs move very slowly along the boundary, a more uniform covering increases the probability that the leak and particle will intersect. In contrast, a non-uniform covering (such as that for $\omega=2/3$) means that the leak is in one of a small number of locations each time the particle strikes the boundary and this gives a lower probability of the particle and leak intersecting. In the case of $\omega=1$, the covering $f_L\left(s\right)$ is $0$ everywhere apart from a small region near $s=0$ and so the leak is, in effect, stationary for a given MUPO. Therefore, the $\alpha$ value for the algebraic tail of $P(t)$ is comparable to those shown in table \ref{t:1} for the stationary leak.

We note that in \eqref{e:5} the quantity that $f_L\left(s\right)$ depends on is $2r\omega$. In all our results so far we have used $r = 1$ and so the denominator of $\omega$ was the denominator of $2r\omega$. When $r$ is varied we find that the results remain the same as those reported here if $\omega$ is scaled by a factor of $1/r$ and $t$ is also scaled by factor $1/r$ (to account for the constant particle velocity $v=1$). Therefore, for a stadium of varied dimensions it is the quantity $2r\omega$ which determines the behaviour of $P(t)$ for a given $\omega$.

\begin{figure}
\begin{center}%
\includegraphics*[width=0.99\columnwidth]{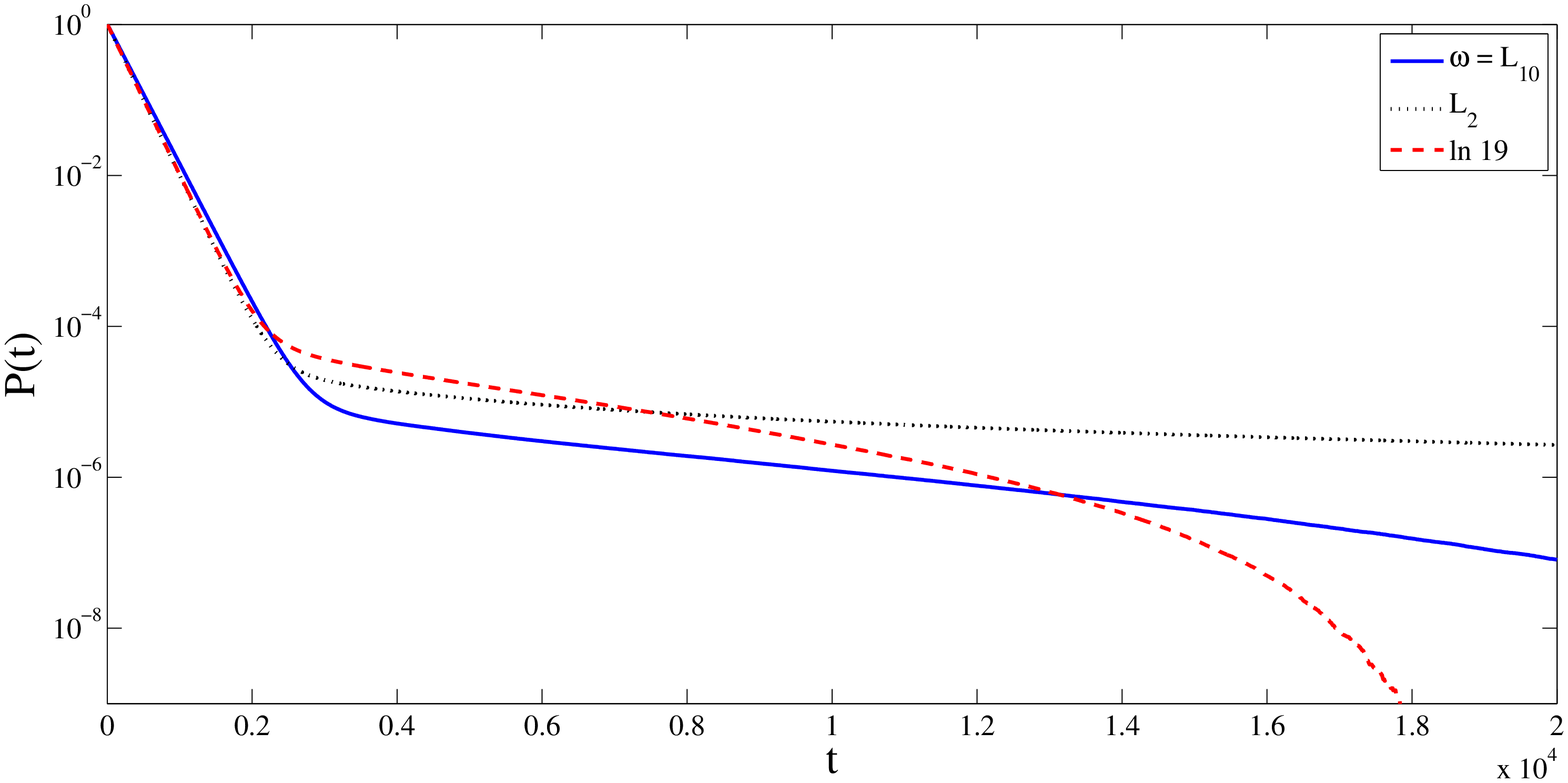}%
\hfil%
\includegraphics*[width=0.99\columnwidth]{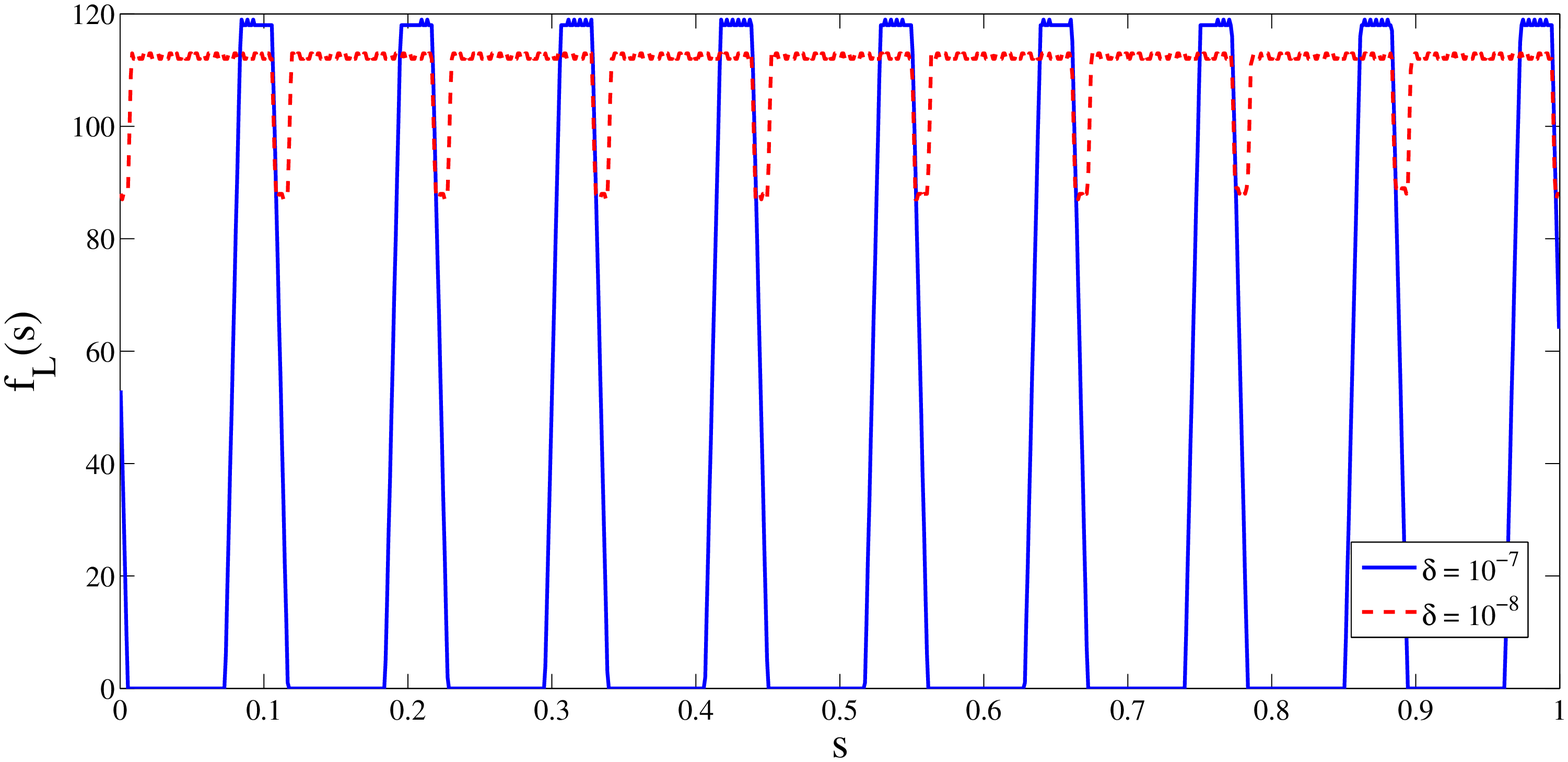}%
\vspace{-1em}
\end{center}
\caption[]{Top: Survival probabilities for three irrational values of $\omega$ that do not show purely exponential decay. Bottom: The covering, $f_L(s)$, for $\omega = \ln\,19$ for two different values of $\delta$.} \label{f:9}
\end{figure}

We conclude this section by briefly considering irrational values of $\omega$. Equation \eqref{e:5} suggests that $f_L(s)$ will be fairly uniform for irrational $\omega$. This is indeed the case for many irrational $\omega$ values and, consequently, $P(t)$ displays purely exponential decay behaviour. However, it is possible to find irrational values of $\omega$ that are sufficiently close to rational $\omega$ values such that $P(t)$ displays an algebraic tail. One such class of irrational numbers is the Liouville numbers\cite{Hardy_numbers} defined as
\begin{equation}
L_{b} = \sum_{k=1}^{\infty} \frac{1}{b^{k!}}
\end{equation}
for any integer $b\geq 2$. The Liouville numbers have the property that for every positive integer $n$, there exists integers $p>0$ and $q>1$ such that
\begin{equation}
0<\left|L_{b}-\frac{p}{q}\right| <\frac{1}{q^n}.
\end{equation}
The Liouville numbers are, therefore, irrational numbers that can be closely approximated by rational numbers. In the top diagram of fig. \ref{f:9} we have plotted survival probability $P(t)$ for $\omega=L_{10}$ and $L_{2}$ which display an algebraic tail. The covering $f_L(s)$ for these numbers contains many intervals of $s$ in which $f_L(s)=0$. We also find irrational numbers other than the Liouville numbers that display non-exponential decay in $P(t)$. For example, in fig. \ref{f:9} we have also plotted $P(t)$ for $\omega = \ln\,19$. The covering for this $\omega$ is interesting as it shows $f_L(s)$ changing from non-uniform (with intervals where $f_L(s)=0$) to uniform as $\delta\rightarrow 0$. This is illustrated in the bottom diagram of fig. \ref{f:9} for two values of $\delta$ and this transition in the covering may explain why the algebraic tail of $P(t)$ is present only at intermdiate times for $\omega=\ln\,19$. Further work is required to establish a more precise relationship between $P(t)$, $f_L(s)$ and $\omega$ for both rational and irrational rotation velocities.

\section{Varied Leak Size}\label{sec:6}
In this work so far we have used a constant leak size of $\lambda = 0.01$. It is known\cite{Altmann_RevModPhys2013} that for a stationary leak, the exponential decay coefficient, $\kappa$, varies approximately with the natural logarithm of $1-\lambda$, whilst the algebraic part of the survival probability depends quadratically on the leak size.\cite{Dettmann_PhysD2009} In this section we investigate the effect of varying $\lambda$ on the three regimes of $\omega$ described in the previous section.

\begin{figure}
\begin{center}%
\includegraphics*[width=0.99\columnwidth]{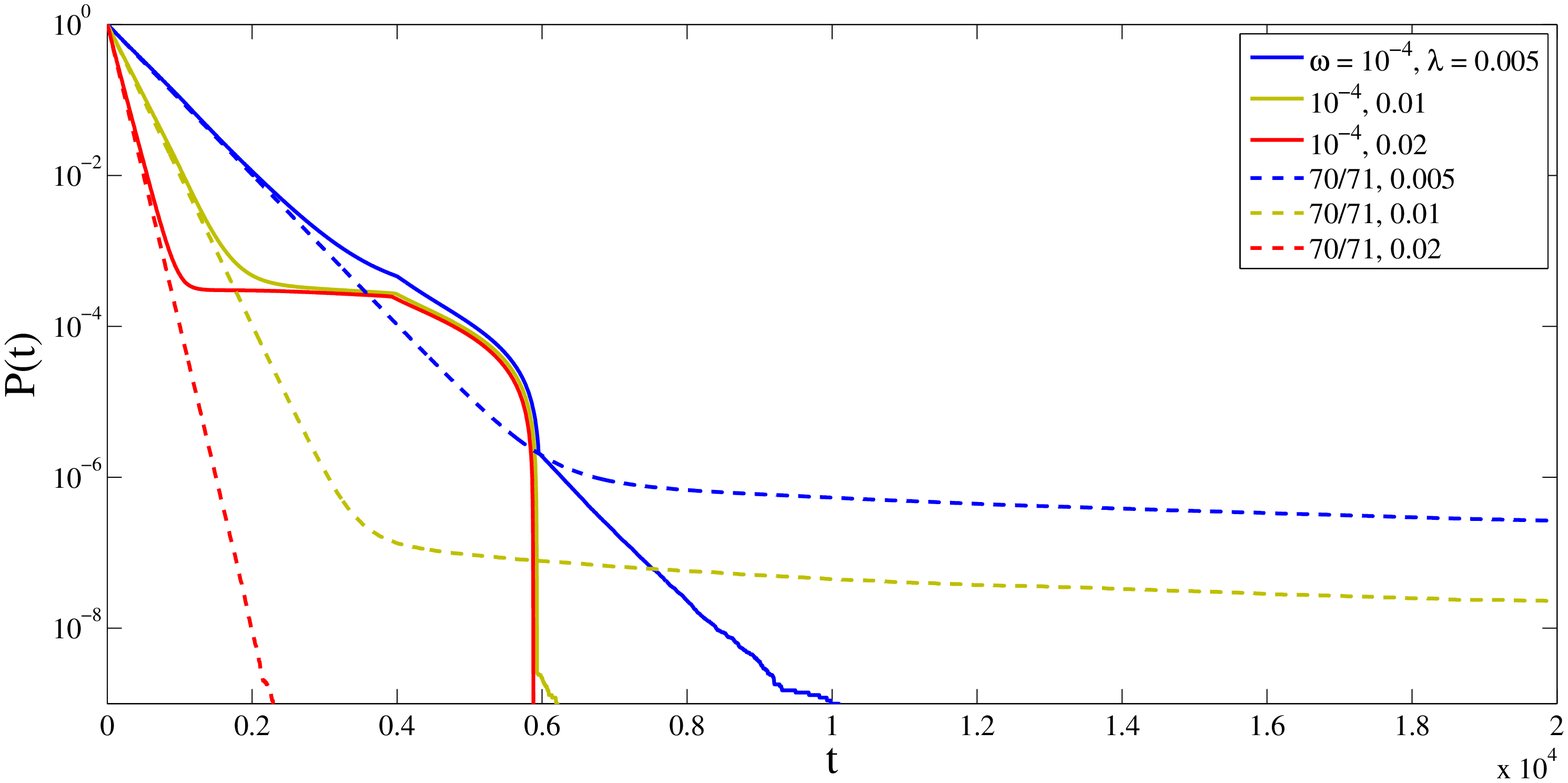}%
\vspace{-1em}
\end{center}
\caption[]{Some examples of the effect of varied leak size on $P(t)$ for $\omega$ values in the low-$\omega$ and high-$\omega$ regime. The exponential decay coefficients are $\kappa=2.25\times10^{-3}, 4.49\times10^{-3}$ and $8.88\times10^{-3}$ for $\lambda=0.005,0.01$ and $0.02$, respectively.} \label{f:8}
\end{figure}

Figure \ref{f:8} shows survival probabilities for a rotation velocity in the low-$\omega$ regime ($\omega=1\times10^{-4}$) and in the high-$\omega$ regime ($\omega=70/71$) for leak sizes of $\lambda = 0.005, 0.01$ and $0.02$. The effect of the leak size on the exponential decay portion of $P(t)$ is clear with a larger leak causing an increase in $\kappa$. 

 Starting with the low-$\omega$ case, the leak size has little effect on the behaviour after the exponential decay phase. This is still dominated by a rapid emptying of the billiard as the leak moves along the lower straight boundary. However, the $\omega$ value at which we move from low-$\omega$ to mid-$\omega$ behaviour does depend on leak size, namely, increasing $\lambda$ increases the $\omega$ value at which this transition occurs. For $\lambda = 0.005$ and $0.02$ we find transition values of $\omega= 1.6\times10^{-4}$ and $5.9 \times10^{-4}$, respectively.

The transition value from the mid-$\omega$ to the high-$\omega$ regime also depends on $\lambda$. For $\lambda = 0.005$ we find the transition occurs at $\omega \approx 3\times10^{-3}$. However, the transition values for $\lambda = 0.01$ and $0.02$ are similar, at $\omega\approx 1\times 10^{-2}$.

Figure \ref{f:8} also shows $P(t)$ for $\omega = 70/71$ for three different leak sizes. As we can see here, the $\alpha$ value of the algebraic tail varies inversely with $\lambda$ (for $\lambda=0.02$ we have not simulated enough particles to accurately measure the $\alpha$ value). These results are consistent with the concept of covering that we introduced in section \ref{sec:4.3}. When we calculate the covering $f_L\left(s\right)$ for $\lambda=0.005$ and $0.01$ we find that there are intervals of $s$ where $f_L\left(s\right)=0$. However, for $\lambda = 0.02$, we find $f_L\left(s\right)>0\quad\forall\quad s\,\,\epsilon\,\,\left[0,1\right]$.

We conclude that with a varied leak size there are still three distinct regimes of $P(t)$ behaviour similar to those described in section \ref{sec:4}. However, the values of $\omega$ in which we transition from one regime to another depend on $\lambda$ and for the high-$\omega$ regime, the $\alpha$ value of the algebraic tail also depends on $\lambda$.

\section{Summary}\label{sec:5}
This paper describes a detailed study of the open stadium billiard with a leak that moves around the boundary at constant velocity. We have identified an interesting relationship between the survival probability $P(t)$ of particles in the stadium and the leak rotational velocity $\omega$. We can summarize the different responses of $P(t)$ to $\omega$ as follows (where the ranges of $\omega$ values are those for the parameters $d=2$, $r=1$, $\lambda=0.01$).
\begin{itemize}
\item[1.] Stationary leak: $\omega = 0$. $P(t)$ exhibits exponential decay at early times but for long times the decay becomes algebraic.
\item[2.] Low-$\omega$ behaviour: $0< \omega< \sim4\times 10^{-4}$. The algebraic tail forms at intermediate times but is then eliminated as the moving leak sweeps up the MUPOs.
\item[3.] Mid-$\omega$ behaviour: $\sim 4\times 10^{-4}\leq\omega < \sim0.01$. The moving leak sweeps up the MUPOs sufficiently quickly such that no algebraic tail is formed. The decay of $P(t)$ is purely exponential.
\item[4.] High-$\omega$ behaviour: $\omega \geq \sim0.01$. The presence and strength ($\alpha$ value) of an algebraic tail in $P(t)$ depends on the quantity $2r\omega$. If this quantity is sufficiently close to a rational number with a low denominator then an algebraic tail is present.
\end{itemize}
For the low-$\omega$ case the behaviour of $P(t)$ evolves during the first cycle of the leak around the stadium boundary while for the mid-$\omega$ case only a few cycles are required to sweep up the MUPOs. However, in the high-$\omega$ case the timescales on which $P(t)$ evolves correspond to multiple cycles of the leak. We also note that even though $\omega$ has a significant effect on the duration of the exponential decay phase, the exponential decay rate remains fairly stable. For all simulations we found $4.3\times10^{-3}\leq\kappa\leq 4.8\times10^{-3}$ for $\lambda = 0.01$.

Our results on varied leak size, described in section \ref{sec:6} indicate that the characteristic behaviour of $P(t)$ in the different $\omega$ regimes is insensitive to leak size, although the transition values of $\omega$ between regimes does depend on $\lambda$.

Our analysis of the results throughout this paper has focussed on the role played by MUPOs. Although the stadium billiard contains rich dynamics and many other classes of orbits, we have not identified any other class of orbits whose survival probability depends on $\omega$ in a similar way to the MUPOs.

Further work is required to establish a more rigorously quantified relationship between $P(t)$ and $\omega$, particularly for the high-$\omega$ regime. However, the results presented here clearly demonstrate that the open stadium billiard with a rotating leak contains interesting and surprising behaviour. We conclude by mentioning that there are many outstanding problems in open billiards\cite{Dettmann_book2011}
that the rotating leak scenario can help to address. These include questions on how to maximize or minimize escape rates and the universality of escape rates.

\begin{acknowledgments}
The author would like to thank Carl Dettmann, Orestis Georgiou and Eduardo Altmann for providing very helpful comments. The results reported in this paper were obtained using the Imperial College supercomputer CX1.
\end{acknowledgments}

\nocite{*}
\bibliography{RotatingLeaks_bib}

\begin{thebibliography}{21}%
\makeatletter
\providecommand \@ifxundefined [1]{%
 \@ifx{#1\undefined}
}%
\providecommand \@ifnum [1]{%
 \ifnum #1\expandafter \@firstoftwo
 \else \expandafter \@secondoftwo
 \fi
}%
\providecommand \@ifx [1]{%
 \ifx #1\expandafter \@firstoftwo
 \else \expandafter \@secondoftwo
 \fi
}%
\providecommand \natexlab [1]{#1}%
\providecommand \enquote  [1]{``#1''}%
\providecommand \bibnamefont  [1]{#1}%
\providecommand \bibfnamefont [1]{#1}%
\providecommand \citenamefont [1]{#1}%
\providecommand \href@noop [0]{\@secondoftwo}%
\providecommand \href [0]{\begingroup \@sanitize@url \@href}%
\providecommand \@href[1]{\@@startlink{#1}\@@href}%
\providecommand \@@href[1]{\endgroup#1\@@endlink}%
\providecommand \@sanitize@url [0]{\catcode `\\12\catcode `\$12\catcode
  `\&12\catcode `\#12\catcode `\^12\catcode `\_12\catcode `\%12\relax}%
\providecommand \@@startlink[1]{}%
\providecommand \@@endlink[0]{}%
\providecommand \url  [0]{\begingroup\@sanitize@url \@url }%
\providecommand \@url [1]{\endgroup\@href {#1}{\urlprefix }}%
\providecommand \urlprefix  [0]{URL }%
\providecommand \Eprint [0]{\href }%
\providecommand \doibase [0]{http://dx.doi.org/}%
\providecommand \selectlanguage [0]{\@gobble}%
\providecommand \bibinfo  [0]{\@secondoftwo}%
\providecommand \bibfield  [0]{\@secondoftwo}%
\providecommand \translation [1]{[#1]}%
\providecommand \BibitemOpen [0]{}%
\providecommand \bibitemStop [0]{}%
\providecommand \bibitemNoStop [0]{.\EOS\space}%
\providecommand \EOS [0]{\spacefactor3000\relax}%
\providecommand \BibitemShut  [1]{\csname bibitem#1\endcsname}%
\let\auto@bib@innerbib\@empty
\bibitem [{\citenamefont {Chernov}\ and\ \citenamefont
  {Markarian}(2006)}]{Chernov_2006}%
  \BibitemOpen
  \bibfield  {author} {\bibinfo {author} {\bibfnamefont {N.}~\bibnamefont
  {Chernov}}\ and\ \bibinfo {author} {\bibfnamefont {R.}~\bibnamefont
  {Markarian}},\ }\href@noop {} {\emph {\bibinfo {title} {Chaotic Billiards,
  Mathematical Surveys and Monographs}}},\ Vol.\ \bibinfo {volume} {127}\
  (\bibinfo  {publisher} {AMS},\ \bibinfo {year} {2006})\BibitemShut {NoStop}%
\bibitem [{\citenamefont {Lai}\ and\ \citenamefont {T\'el}(2011)}]{Tel_2011}%
  \BibitemOpen
  \bibfield  {author} {\bibinfo {author} {\bibfnamefont {Y.-C.}\ \bibnamefont
  {Lai}}\ and\ \bibinfo {author} {\bibfnamefont {T.}~\bibnamefont {T\'el}},\
  }\href@noop {} {\emph {\bibinfo {title} {Transient Chaos, Complex Dynamics on
  Finite-Time Scales}}}\ (\bibinfo  {publisher} {Springer, New York},\ \bibinfo
  {year} {2011})\BibitemShut {NoStop}%
\bibitem [{\citenamefont {T\'el}(2015)}]{Tel_Chaos2015}%
  \BibitemOpen
  \bibfield  {author} {\bibinfo {author} {\bibfnamefont {T.}~\bibnamefont
  {T\'el}},\ }\bibfield  {title} {\enquote {\bibinfo {title} {The joy of
  transient chaos},}\ }\href {\doibase http://dx.doi.org/10.1063/1.4917287}
  {\bibfield  {journal} {\bibinfo  {journal} {Chaos}\ }\textbf {\bibinfo
  {volume} {25}},\ \bibinfo {eid} {097619} (\bibinfo {year} {2015}),\
  http://dx.doi.org/10.1063/1.4917287}\BibitemShut {NoStop}%
\bibitem [{\citenamefont {Altmann}, \citenamefont {Portela},\ and\
  \citenamefont {T\'el}(2013)}]{Altmann_RevModPhys2013}%
  \BibitemOpen
  \bibfield  {author} {\bibinfo {author} {\bibfnamefont {E.~G.}\ \bibnamefont
  {Altmann}}, \bibinfo {author} {\bibfnamefont {J.~S.~E.}\ \bibnamefont
  {Portela}}, \ and\ \bibinfo {author} {\bibfnamefont {T.}~\bibnamefont
  {T\'el}},\ }\bibfield  {title} {\enquote {\bibinfo {title} {Leaking chaotic
  systems},}\ }\href {\doibase 10.1103/RevModPhys.85.869} {\bibfield  {journal}
  {\bibinfo  {journal} {Rev. Mod. Phys.}\ }\textbf {\bibinfo {volume} {85}},\
  \bibinfo {pages} {869--918} (\bibinfo {year} {2013})}\BibitemShut {NoStop}%
\bibitem [{\citenamefont {Bunimovich}(1974)}]{Bunimovich_1974}%
  \BibitemOpen
  \bibfield  {author} {\bibinfo {author} {\bibfnamefont {L.~A.}\ \bibnamefont
  {Bunimovich}},\ }\bibfield  {title} {\enquote {\bibinfo {title} {On ergodic
  properties of certain billiards},}\ }\href {\doibase 10.1007/BF01075700}
  {\bibfield  {journal} {\bibinfo  {journal} {Functional Analysis and Its
  Applications}\ }\textbf {\bibinfo {volume} {8}},\ \bibinfo {pages} {254--255}
  (\bibinfo {year} {1974})}\BibitemShut {NoStop}%
\bibitem [{\citenamefont {Bunimovich}(1979)}]{Bunimovich_1979}%
  \BibitemOpen
  \bibfield  {author} {\bibinfo {author} {\bibfnamefont {L.~A.}\ \bibnamefont
  {Bunimovich}},\ }\bibfield  {title} {\enquote {\bibinfo {title} {On the
  ergodic properties of nowhere dispersing billiards},}\ }\href {\doibase
  10.1007/BF01197884} {\bibfield  {journal} {\bibinfo  {journal}
  {Communications in Mathematical Physics}\ }\textbf {\bibinfo {volume} {65}},\
  \bibinfo {pages} {295--312} (\bibinfo {year} {1979})}\BibitemShut {NoStop}%
\bibitem [{\citenamefont {Alt}\ \emph {et~al.}(1996)\citenamefont {Alt},
  \citenamefont {Gr\"af}, \citenamefont {Harney}, \citenamefont {Hofferbert},
  \citenamefont {Rehfeld}, \citenamefont {Richter},\ and\ \citenamefont
  {Schardt}}]{Alt_PRE1995}%
  \BibitemOpen
  \bibfield  {author} {\bibinfo {author} {\bibfnamefont {H.}~\bibnamefont
  {Alt}}, \bibinfo {author} {\bibfnamefont {H.-D.}\ \bibnamefont {Gr\"af}},
  \bibinfo {author} {\bibfnamefont {H.~L.}\ \bibnamefont {Harney}}, \bibinfo
  {author} {\bibfnamefont {R.}~\bibnamefont {Hofferbert}}, \bibinfo {author}
  {\bibfnamefont {H.}~\bibnamefont {Rehfeld}}, \bibinfo {author} {\bibfnamefont
  {A.}~\bibnamefont {Richter}}, \ and\ \bibinfo {author} {\bibfnamefont
  {P.}~\bibnamefont {Schardt}},\ }\bibfield  {title} {\enquote {\bibinfo
  {title} {Decay of classical chaotic systems: The case of the bunimovich
  stadium},}\ }\href {\doibase 10.1103/PhysRevE.53.2217} {\bibfield  {journal}
  {\bibinfo  {journal} {Phys. Rev. E}\ }\textbf {\bibinfo {volume} {53}},\
  \bibinfo {pages} {2217--2222} (\bibinfo {year} {1996})}\BibitemShut {NoStop}%
\bibitem [{\citenamefont {Dumont}\ and\ \citenamefont
  {Brumer}(1992)}]{DUMONT_CPL1992}%
  \BibitemOpen
  \bibfield  {author} {\bibinfo {author} {\bibfnamefont {R.~S.}\ \bibnamefont
  {Dumont}}\ and\ \bibinfo {author} {\bibfnamefont {P.}~\bibnamefont
  {Brumer}},\ }\bibfield  {title} {\enquote {\bibinfo {title} {Decay of a
  chaotic dynamical system},}\ }\href {\doibase
  http://dx.doi.org/10.1016/0009-2614(92)80867-B} {\bibfield  {journal}
  {\bibinfo  {journal} {Chemical Physics Letters}\ }\textbf {\bibinfo {volume}
  {188}},\ \bibinfo {pages} {565 -- 571} (\bibinfo {year} {1992})}\BibitemShut
  {NoStop}%
\bibitem [{\citenamefont {Armstead}, \citenamefont {Hunt},\ and\ \citenamefont
  {Ott}(2004)}]{Armstead_PhysD2004}%
  \BibitemOpen
  \bibfield  {author} {\bibinfo {author} {\bibfnamefont {D.~N.}\ \bibnamefont
  {Armstead}}, \bibinfo {author} {\bibfnamefont {B.~R.}\ \bibnamefont {Hunt}},
  \ and\ \bibinfo {author} {\bibfnamefont {E.}~\bibnamefont {Ott}},\ }\bibfield
   {title} {\enquote {\bibinfo {title} {Power-law decay and self-similar
  distributions in stadium-type billiards},}\ }\href {\doibase
  http://dx.doi.org/10.1016/j.physd.2004.01.013} {\bibfield  {journal}
  {\bibinfo  {journal} {Physica D: Nonlinear Phenomena}\ }\textbf {\bibinfo
  {volume} {193}},\ \bibinfo {pages} {96 -- 127} (\bibinfo {year} {2004})},\
  \bibinfo {note} {anomalous distributions, nonlinear dynamics, and
  nonextensivity}\BibitemShut {NoStop}%
\bibitem [{\citenamefont {Bunimovich}\ and\ \citenamefont
  {Vela-Arevalo}(2012)}]{Bunimovich_Chaos2012}%
  \BibitemOpen
  \bibfield  {author} {\bibinfo {author} {\bibfnamefont {L.~A.}\ \bibnamefont
  {Bunimovich}}\ and\ \bibinfo {author} {\bibfnamefont {L.~V.}\ \bibnamefont
  {Vela-Arevalo}},\ }\bibfield  {title} {\enquote {\bibinfo {title} {Many faces
  of stickiness in hamiltonian systems},}\ }\href {\doibase
  http://dx.doi.org/10.1063/1.3692974} {\bibfield  {journal} {\bibinfo
  {journal} {Chaos}\ }\textbf {\bibinfo {volume} {22}},\ \bibinfo {eid}
  {026103} (\bibinfo {year} {2012}),\
  http://dx.doi.org/10.1063/1.3692974}\BibitemShut {NoStop}%
\bibitem [{\citenamefont {{Dettmann}}(2016)}]{Dettmann_arXiv2016}%
  \BibitemOpen
  \bibfield  {author} {\bibinfo {author} {\bibfnamefont {C.~P.}\ \bibnamefont
  {{Dettmann}}},\ }\bibfield  {title} {\enquote {\bibinfo {title} {{How sticky
  is the chaos/order boundary?}}}\ }\href@noop {} {\bibfield  {journal}
  {\bibinfo  {journal} {ArXiv e-prints}\ } (\bibinfo {year} {2016})},\ \Eprint
  {http://arxiv.org/abs/1603.00667} {arXiv:1603.00667 [nlin.CD]} \BibitemShut
  {NoStop}%
\bibitem [{\citenamefont {Dettmann}\ and\ \citenamefont
  {Georgiou}(2009)}]{Dettmann_PhysD2009}%
  \BibitemOpen
  \bibfield  {author} {\bibinfo {author} {\bibfnamefont {C.~P.}\ \bibnamefont
  {Dettmann}}\ and\ \bibinfo {author} {\bibfnamefont {O.}~\bibnamefont
  {Georgiou}},\ }\bibfield  {title} {\enquote {\bibinfo {title} {Survival
  probability for the stadium billiard},}\ }\href {\doibase
  http://dx.doi.org/10.1016/j.physd.2009.09.019} {\bibfield  {journal}
  {\bibinfo  {journal} {Physica D: Nonlinear Phenomena}\ }\textbf {\bibinfo
  {volume} {238}},\ \bibinfo {pages} {2395 -- 2403} (\bibinfo {year}
  {2009})}\BibitemShut {NoStop}%
\bibitem [{\citenamefont {Altmann}, \citenamefont {Leitão},\ and\
  \citenamefont {Viana~Lopes}(2012)}]{Altmann_Chaos2012}%
  \BibitemOpen
  \bibfield  {author} {\bibinfo {author} {\bibfnamefont {E.~G.}\ \bibnamefont
  {Altmann}}, \bibinfo {author} {\bibfnamefont {J.~C.}\ \bibnamefont
  {Leitão}}, \ and\ \bibinfo {author} {\bibfnamefont {J.}~\bibnamefont
  {Viana~Lopes}},\ }\bibfield  {title} {\enquote {\bibinfo {title} {Effect of
  noise in open chaotic billiards},}\ }\href {\doibase
  http://dx.doi.org/10.1063/1.3697408} {\bibfield  {journal} {\bibinfo
  {journal} {Chaos}\ }\textbf {\bibinfo {volume} {22}},\ \bibinfo {eid}
  {026114} (\bibinfo {year} {2012}),\
  http://dx.doi.org/10.1063/1.3697408}\BibitemShut {NoStop}%
\bibitem [{\citenamefont {Bunimovich}\ and\ \citenamefont
  {Dettmann}(2005)}]{Bunimovich_PRL2005}%
  \BibitemOpen
  \bibfield  {author} {\bibinfo {author} {\bibfnamefont {L.~A.}\ \bibnamefont
  {Bunimovich}}\ and\ \bibinfo {author} {\bibfnamefont {C.~P.}\ \bibnamefont
  {Dettmann}},\ }\bibfield  {title} {\enquote {\bibinfo {title} {Open circular
  billiards and the riemann hypothesis},}\ }\href {\doibase
  10.1103/PhysRevLett.94.100201} {\bibfield  {journal} {\bibinfo  {journal}
  {Phys. Rev. Lett.}\ }\textbf {\bibinfo {volume} {94}},\ \bibinfo {pages}
  {100201} (\bibinfo {year} {2005})}\BibitemShut {NoStop}%
\bibitem [{\citenamefont {Dettmann}\ and\ \citenamefont
  {Georgiou}(2011)}]{Dettmann_PRE2011}%
  \BibitemOpen
  \bibfield  {author} {\bibinfo {author} {\bibfnamefont {C.~P.}\ \bibnamefont
  {Dettmann}}\ and\ \bibinfo {author} {\bibfnamefont {O.}~\bibnamefont
  {Georgiou}},\ }\bibfield  {title} {\enquote {\bibinfo {title} {Transmission
  and reflection in the stadium billiard: Time-dependent asymmetric
  transport},}\ }\href {\doibase 10.1103/PhysRevE.83.036212} {\bibfield
  {journal} {\bibinfo  {journal} {Phys. Rev. E}\ }\textbf {\bibinfo {volume}
  {83}},\ \bibinfo {pages} {036212} (\bibinfo {year} {2011})}\BibitemShut
  {NoStop}%
\bibitem [{\citenamefont {Leonel}\ and\ \citenamefont
  {Dettmann}(2012)}]{Leonel_PLA2012}%
  \BibitemOpen
  \bibfield  {author} {\bibinfo {author} {\bibfnamefont {E.~D.}\ \bibnamefont
  {Leonel}}\ and\ \bibinfo {author} {\bibfnamefont {C.~P.}\ \bibnamefont
  {Dettmann}},\ }\bibfield  {title} {\enquote {\bibinfo {title} {Recurrence of
  particles in static and time varying oval billiards},}\ }\href {\doibase
  http://dx.doi.org/10.1016/j.physleta.2012.03.056} {\bibfield  {journal}
  {\bibinfo  {journal} {Physics Letters A}\ }\textbf {\bibinfo {volume}
  {376}},\ \bibinfo {pages} {1669 -- 1674} (\bibinfo {year}
  {2012})}\BibitemShut {NoStop}%
\bibitem [{\citenamefont {Bunimovich}\ and\ \citenamefont
  {Dettmann}(2007)}]{Bunimovich_EPL2007}%
  \BibitemOpen
  \bibfield  {author} {\bibinfo {author} {\bibfnamefont {L.~A.}\ \bibnamefont
  {Bunimovich}}\ and\ \bibinfo {author} {\bibfnamefont {C.~P.}\ \bibnamefont
  {Dettmann}},\ }\bibfield  {title} {\enquote {\bibinfo {title} {Peeping at
  chaos: Nondestructive monitoring of chaotic systems by measuring long-time
  escape rates},}\ }\href {http://stacks.iop.org/0295-5075/80/i=4/a=40001}
  {\bibfield  {journal} {\bibinfo  {journal} {EPL (Europhysics Letters)}\
  }\textbf {\bibinfo {volume} {80}},\ \bibinfo {pages} {40001} (\bibinfo {year}
  {2007})}\BibitemShut {NoStop}%
\bibitem [{\citenamefont {{Hansen}}, \citenamefont {{Egydio de Carvalho}},\
  and\ \citenamefont {{Leonel}}(2016)}]{Hansen_arXiv2016}%
  \BibitemOpen
  \bibfield  {author} {\bibinfo {author} {\bibfnamefont {M.}~\bibnamefont
  {{Hansen}}}, \bibinfo {author} {\bibfnamefont {R.}~\bibnamefont {{Egydio de
  Carvalho}}}, \ and\ \bibinfo {author} {\bibfnamefont {E.~D.}\ \bibnamefont
  {{Leonel}}},\ }\bibfield  {title} {\enquote {\bibinfo {title} {{Influence of
  stability islands in the recurrence of particles in a static oval billiard
  with holes}},}\ }\href@noop {} {\bibfield  {journal} {\bibinfo  {journal}
  {ArXiv e-prints}\ } (\bibinfo {year} {2016})},\ \Eprint
  {http://arxiv.org/abs/1602.08353} {arXiv:1602.08353 [nlin.CD]} \BibitemShut
  {NoStop}%
\bibitem [{\citenamefont {Dettmann}\ and\ \citenamefont
  {Georgiou}(2012)}]{Dettmann_Chaos2012}%
  \BibitemOpen
  \bibfield  {author} {\bibinfo {author} {\bibfnamefont {C.~P.}\ \bibnamefont
  {Dettmann}}\ and\ \bibinfo {author} {\bibfnamefont {O.}~\bibnamefont
  {Georgiou}},\ }\bibfield  {title} {\enquote {\bibinfo {title} {Quantifying
  intermittency in the open drivebelt billiard},}\ }\href {\doibase
  http://dx.doi.org/10.1063/1.3685522} {\bibfield  {journal} {\bibinfo
  {journal} {Chaos}\ }\textbf {\bibinfo {volume} {22}},\ \bibinfo {eid}
  {026113} (\bibinfo {year} {2012}),\
  http://dx.doi.org/10.1063/1.3685522}\BibitemShut {NoStop}%
\bibitem [{\citenamefont {Hardy}\ and\ \citenamefont
  {Wright}(2008)}]{Hardy_numbers}%
  \BibitemOpen
  \bibfield  {author} {\bibinfo {author} {\bibfnamefont {G.~H.}\ \bibnamefont
  {Hardy}}\ and\ \bibinfo {author} {\bibfnamefont {E.~M.}\ \bibnamefont
  {Wright}},\ }\href@noop {} {\emph {\bibinfo {title} {An Introduction to the
  Theory of Numbers}}},\ \bibinfo {edition} {6th}\ ed.\ (\bibinfo  {publisher}
  {Oxford University Press},\ \bibinfo {year} {2008})\BibitemShut {NoStop}%
\bibitem [{\citenamefont {Dettmann}(2011)}]{Dettmann_book2011}%
  \BibitemOpen
  \bibfield  {author} {\bibinfo {author} {\bibfnamefont {C.~P.}\ \bibnamefont
  {Dettmann}},\ }\enquote {\bibinfo {title} {Frontiers in the study of chaotic
  dynamical systems with open problems},}\ \ (\bibinfo  {publisher} {World
  Scientific},\ \bibinfo {year} {2011})\ Chap.\ \bibinfo {chapter} {Recent
  Advances in Open Billiards with Some Open Problems}\BibitemShut {NoStop}%
\end{thebibliography}%



\end{document}